\documentclass[11pt,epsf]{article}
\usepackage{amsmath}
\usepackage{amsfonts}
\usepackage{amssymb}
\usepackage{graphicx}

\newcommand {\dd} {\mbox{d}}
\newcommand {\dfn} {\stackrel{\Delta} {=}}
\newcommand {\exe} {\stackrel{\cdot} {=}}

\newcommand {\reals} {{\rm I\!R}}

\newcommand {\bn} {\mbox{\boldmath $n$}}

\newcommand {\br} {\mbox{\boldmath $r$}}
\newcommand {\bs} {\mbox{\boldmath $s$}}

\newcommand {\bE} {\mbox{\boldmath $E$}}

\newcommand{\calH}{{\cal H}}
\newcommand{\calI}{{\cal I}}

\newcommand{\calM}{{\cal M}}
\newcommand{\calN}{{\cal N}}

\begin{document}
\thispagestyle{empty}
\title{Threshold Effects in Parameter Estimation as Phase Transitions in
Statistical Mechanics}
\author{Neri Merhav}
\date{}
\maketitle

\begin{center}
Department of Electrical Engineering \\
Technion - Israel Institute of Technology \\
Haifa 32000, ISRAEL \\
{\tt merhav@ee.technion.ac.il}
\end{center}
\vspace{1.5\baselineskip}
\setlength{\baselineskip}{1.5\baselineskip}

\begin{abstract}
Threshold effects in the estimation of parameters of non--linearly
modulated, continuous--time, wide-band waveforms,
are examined from a statistical physics perspective. These threshold effects
are shown to be analogous to phase transitions of certain disordered physical
systems in thermal equilibrium. The main message, in this work, is in 
demonstrating that this physical point
of view may be insightful for understanding the interactions between two or
more parameters to be estimated, from the aspects of the threshold effect.\\


\end{abstract}

\newpage
\section{Introduction}

In waveform communication systems, the information is normally conveyed in a
real--valued parameter (or parameters) of a continuous--time signal
to be transmitted, whereas the receiver is based on
estimating this parameter from a noisy received 
version of this signal \cite[Chap.\ 8]{WJ65}. 
This concept of mapping a real--valued parameter, or a parameter vector, into a
continuous--time signal, using a certain modulation scheme, stands at the
basis of the theory and practice of Shannon--Kotel'nikov mappings,
which can in turn be viewed as certain families of joint source--channel codes
(see, e.g., \cite{Floor08},\cite{FR09},\cite{Hekland07},\cite{Ramstad02} 
as well as many references therein).

When the underlying modulation scheme is highly non--linear, like in
frequency modulation (FM), phase modulation (PM), pulse position modulation
(PPM), or frequency position modulation (FPM), it is well known that
the estimation of the desired parameter is subjected to a {\it threshold
effect}. This threshold effect means that the wider is the bandwidth of the transmitted
signal, the better is the accuracy of the maximum likelihood (ML) estimator at the high signal--to--noise
ratio (SNR) regime, but on the other hand, it 
comes at the price of increasing also a certain critical level of
the SNR, referred to as the
{\it threshold SNR}, below which this estimator breaks down. 
This breakdown means that the estimator makes
gross errors (a.k.a.\ anomalous errors) with an overwhelmingly large
probability, and in the high bandwidth regime, this breakdown becomes abrupt,
as the SNR crosses the threshold value.
This threshold effect
is not merely an artifact to be attributed to a specific modulator and/or
estimation method. It is a fundamental limitation which is inherent
to any (non--linear) communication system operating under a limited power
constraint over a wide-band channel.

In this paper, we propose a statistical--mechanical perspective on the
threshold effect. According to this perspective, the abrupt threshold effect
of the wide-band regime is viewed as a {\it phase transition} of a certain
disordered physical system of interacting particles. Specifically, this
physical system turns out to be closely related 
(though not quite identical) to a well--known model
in the statistical physics literature, which is called the
{\it random energy model} (REM). 
The REM is one model (among many other models) for highly disordered
magnetic materials, called {\it spin glasses}. The REM
was invented by Derrida in the early
eighties of the previous century
\cite{Derrida80},\cite{Derrida80b},\cite{Derrida81}, and it was shown more recently in
\cite[Chap.\ 6]{MM09} (see also \cite{Merhav08}) 
to be intimately related to 
phase transitions in the behavior of
ensembles of random channel codes, not merely in the
context of ordinary digital decoding, but also in minimum mean square error (MMSE)
signal estimation \cite{MGS10}. 

This paper, in contrast to \cite{MGS10},
examines the physics of the threshold effect in the estimation of
a continuous--valued parameter, rather than the estimation of the signal itself.
For the sake of simplicity and concreteness, the analogy between 
the threshold effect and phase transitions
is demonstrated in the context of estimating the delay (or the position) of a
narrow rectangular pulse, but the methodology is generalizable to other situations, as
discussed in the sequel. A phase diagram with three phases 
(similarly as in \cite{MM09}) is obtained in the plane of two
design parameters of the communication system, one pertaining to the signal bandwidth, and
the other to a certain notion of temperature (which will be made clear in the
sequel). 

Beyond the fact that this relationship, between the threshold 
effect in parameter estimation and phase transitions in physics,
may be interesting on its own right, we also believe that the physical point
of view may provide insights and tools for understanding the interactions
and the collective behavior of the 
joint ML estimators of two or more parameters,
in the context of the threshold effect. For example, suppose that both the
amplitude and the delay of a narrow pulse are to be estimated. While the
amplitude estimation alone does not exhibit any threshold effect (as the
modulation is linear) and the delay estimation alone displays a
phase diagram with three phases, it turns out that when 
joint ML estimation of both amplitude and delay is considered, the interaction
between them exhibits a surprisingly more 
erratic behavior, than
that of the delay parameter alone: It
possesses as many as five different phases in the plane of
bandwidth vs.\ temperature. Moreover, the behavior of the anomalous errors
(below the threshold) pertaining to the 
amplitude and the delay are very different in
character, and it is the physical point of view that gives rise to
understanding them. 

The outline of this paper is as follows. In Section \ref{bcgd}, we provide
some basic background on the threshold effect in non--linear modulation and
estimation. In Section \ref{phys}, we present the threshold effect from the
physics viewpoint and, in particular, we show 
how it is related to phase transitions pertaining
to the REM. In Section \ref{joint}, we consider joint ML estimation
of amplitude and delay, as described in the previous paragraph, and provide
the phase diagram. Finally, in Section \ref{conc}, we summarize and conclude
this work.

\section{Background}
\label{bcgd}

We begin with some basic background on 
ML parameter estimation for non--linearly
modulated signals in additive white Gaussian noise (AWGN), the threshold
effect pertaining to this estimation, and 
then the signal design problem, first, for band--limited signals, and
then in large bandwidth limit. The material in this section, which is mostly
classical and can be found in \cite[Chap.\ 8]{WJ65}, is briefly reviewed here
merely for the sake of completeness and convenience of the reader.

Consider the following estimation problem. 
We are given a parametric family of
waveforms $\{s_m(t),~-T/2 \le t\le +T/2\}$, 
where $m$ is the parameter, which for
convenience, will be assumed a (deterministic) scalar that takes on values in
some interval
$[-M,+M]$, ($M>0$). Now suppose 
that we observe a noisy version of $s_m(t)$ along
the time interval $[-T/2,+T/2]$, i.e.,
\begin{equation}
r(t)=s_m(t)+n(t),~~~-\frac{T}{2}\le t\le +\frac{T}{2}
\end{equation}
where $\{n(t)\}$ is a zero--mean Gaussian white noise with spectral density
$N_0/2$, and we wish to estimate
$m$ from $\br=\{r(t),~-T/2\le t\le +T/2\}$. 
Maximum likelihood (ML) estimation, in
the Gaussian case considered here, 
is obviously equivalent to the minimization of
\begin{equation}
\int_0^T[r(t)-s_m(t)]^2\dd t
\end{equation}
w.r.t.\ $m$. The simplest example is the one where the parametrization of the
signal is linear in $m$, i.e., $s_m(t)=m\cdot s(t)$, where
$\{s(t),~-T/2\le t\le +T/2\}$ is a
given waveform (independent of $m$).
In this case, ML estimation yields
\begin{equation}
\hat{m}=\frac{\int_0^Tr(t)s(t)\dd t}{\int_0^Ts^2(t)\dd t}=
\frac{\int_0^Tr(t)s(t)\dd t}{E},
\end{equation}
where $E$ designates the energy of $\{s(t)\}$, i.e., $E=\int_0^Ts^2(t)\dd
t$, and mean square error (MSE) is readily obtained as
\begin{equation}
\bE\{(\hat{m}-m)^2\}
=\frac{N_0}{2E}.
\end{equation}
The estimation performance depends
on the signal $\{s(t)\}$ only via its energy, $E$.
Since this MSE achieves the Cram\'er--Rao lower bound, this
is essentially the best one can 
do (at least as far as unbiased estimators go) with linear
parametrization, for a given SNR $E/N_0$. 

The only
way then to improve on this result, at least for very large SNR,
is to extend the scope to
non--linear parametrizations of $\{s_m(t)\}$. For example, $m$ can stand for
the {\it delay} (or the {\it position}) of a given pulse $s(t)$, i.e.,
$s_m(t)=s(t-m)$. Also, in the case of a sinusoidal waveform,
$s(t)=A\sin(\omega t+\phi)$ (with $A$, $\omega$ and $\phi$
being fixed parameters), $m$ can designate a frequency offset, as in
$s_m(t)=A\sin[(\omega+m)t+\phi]$, or a phase offset as in
$s_m(t)=A\sin(\omega t+\phi+m)$. In these examples, the MSE in
the high SNR regime, depends not only
on the SNR, $E/N_0$, but also on the shape of the waveform,
i.e., on some notion of bandwidth: Rapidly varying signals
can be estimated more accurately than slowly varying ones.
To demonstrate this, let us assume that the noise is very weak, and 
the true parameter is $m=m_0$. For small deviations from $m_0$, 
we consider the linearization
\begin{equation}
s_m(t)\approx s_{m_0}(t) +(m-m_0)\dot{s}_{m_0}(t),
\end{equation}
where $\dot{s}_{m_0}(t)=\mbox{d}s_m(t)/\mbox{d}m|_{m=m_0}$. 
This is then essentially
the same linear model as before where the previous 
role of $\{s(t)\}$ is played
now by $\{\dot{s}_{m_0}(t)\}$, and so, the MSE is about 
\begin{equation}
\bE\{(\hat{m}-m)^2\}\approx\frac{N_0}{2\dot{E}},
\end{equation}
where
$\dot{E}$ is the energy of $\{\dot{s}_{m_0}(t)\}$, 
which depends, of course, not only on
$E$, but also on the shape of $\{s_m(t)\}$. For example, if $m$
is a delay parameter, $s_m(t)=s(t-m)$, and $\{s(t)\}$ 
contains a narrow pulse (or pulses) compared
to $T$, then $\dot{E}=\int_0^T\dot{s}^2(t)\mbox{d}t$, essentially
independently of $m$, where $\dot{s}(t)$ is the time derivative of $s(t)$.
By the Parseval theorem, 
\begin{equation}
\int_0^T\dot{s}^2(t)\mbox{d}t=
\int_{-\infty}^{+\infty}\mbox{d}f (2\pi f)^2S(f),
\end{equation}
where
$S(f)$ is the Fourier transform of $\{s(t)\}$, 
and so, we have $\dot{E}=W^2E$ where $W$ is
the effective bandwidth of $s(t)$ in the 
second moment sense, a.k.a.\ the {\it
Gabor bandwidth}. We then have
\begin{equation}
\bE\{(\hat{m}-m)^2\}\approx\frac{N_0}{2W^2E},
\end{equation}
which means that MSE depends, not only on $E/N_0$, but also on the signal
shape -- in this case, its Gabor bandwidth, $W$. One might be tempted
to think that the larger is $W$, the better is the MSE. However, there is a
price for increasing $W$: the probability of {\it anomalous errors}
increases.

To understand the effect of anomaly, it is instructive to look at the broader
picture: Let us assume that the parametric family of signals $\{s_m(t):~-M\le
m\le +M\}$ lies in the linear space spanned by a set of $K$ orthonormal basis
functions $\{\phi_i(t)\}_{i=1}^K$, defined over $-T/2\le t\le +T/2$, 
and so, we can
pass from continuous time signals to vectors of coefficients:
\begin{equation}
s(t)=\sum_{i=1}^Ks_i(m)\phi_i(t)
\end{equation}
with
\begin{equation}
s_i(m)=\int_0^Ts(t)\phi_i(t)\mbox{d}t,
\end{equation}
and let us apply similar decompositions to $r(t)$ and $n(t)$, so as to obtain
vectors of coefficients $\br=(r_1,\ldots,r_K)$, and
$\bn=(n_1,\ldots,n_K)$, related by
\begin{equation}
r_i=s_i(m)+n_i,~~~~i=1,2,\ldots,K
\end{equation}
where $n_i\sim\calN(0,N_0/2)$, or 
\begin{equation}
\br=\bs(m)+\bn.
\end{equation}
As in the example of a delay parameter, let us assume that both the energy $E$
of the signal $\{s_m(t)\}$ itself, and 
the energy $\dot{E}$ of its derivative w.r.t.\ $m$,
$\{\dot{s}_m(t)\}$, are fixed, independently of $m$. In other words,
$\sum_i s_i^2(m)=E$ and
$\sum_i \dot{s}_i^2(m)=\dot{E}$ for all $m$. Consider the locus of the signal
vectors $[s_1(m),\ldots,s_K(m)]$ in $\reals^K$
as $m$ varies from $-M$ to $+M$. 
On the one hand, this locus is constrained to lie on the hyper-surface of an
$K$--dimensional sphere of radius $\sqrt{E}$, on the other hand, since
the high--SNR MSE behaves according to $N_0/(2\dot{E})$, we would like
$\dot{E}=\sum_i\dot{s}_i^2(m)$ to be as large as possible. But
$\dot{E}$ is related to the length $L$ 
of the signal locus in $\reals^K$ according to
\begin{equation}
L=\int_{-M}^{+M}\mbox{d}m\sqrt{\sum_i\dot{s}_i^2(m)}=2M\sqrt{\dot{E}},
\end{equation}
where we have used the assumption that the norm of
$\dot{\bs}(m)=(\dot{s}_1(m),\ldots,\dot{s}_K(m))$ is independent of $m$.
Thus, the high--SNR MSE is about 
\begin{equation}
\bE\{(\hat{m}-m)^2\}\approx\frac{2N_0M^2}{L^2},
\end{equation}
which means that we would like to make the signal locus
as long as possible,
in order to minimize the high--SNR MSE. 

Our problem is then to design a signal locus, 
as long as possible, which lies in the hyper-surface of a $K$--dimensional
sphere of radius $\sqrt{E}$. Since our room is limited by this energy
constraint, a long locus would mean that it is very curvy, 
with many sharp foldings, and there must then
be pairs of points $m_1$ and $m_2$, which are 
far apart, yet $\bs(m_1)$ and $\bs(m_2)$ 
are close in the Euclidean distance sense. 
In this case, if the noise
vector $\bn$ has a sufficiently 
large projection in the direction of $\bs(m_2)-\bs(m_1)$, 
it can cause a gross error, confusing $m_1$ with $m_2$.
Moreover, in high
dimension $K$, there can be much more 
than one such problematic (orthogonal) direction 
in the above described sense 
and then the event of anomalous error,
which is the event that the noise projection is large in at least one of these
directions, gains an appreciably large probability.
Thus, as the locus of $\bs(m)$ bends, various folds of the curve
must be kept sufficiently far apart in 
all dimensions, so that the noise cannot cause anomalous
errors with high probability. The 
probability of anomaly then sets the limit on the length of the
curve, and hence also on the high SNR MSE.
The maximum locus length $L$ is 
shown in \cite{WJ65} to grow exponentially 
at the rate of $e^{CT}$ in the large $T$ limit,
where $C$ is the capacity of the infinite--bandwidth
AWGN channel, given by $C=P/N_0$, with $P=E/T$ being the signal power.
This maximum is essentially attained by the family
{\it frequency--position modulation} (FPM) signals (see \cite{WJ65}),
as well as by {\it pulse--position modulation} (PPM) signals, considered
hereafter.

As is shown in \cite[Chap.\ 8]{WJ65}, if the signal space 
is spanned by $K\sim 2WT$ dimensions of 
signals of duration $T$ and fixed bandwidth
$W$, namely, $K$ grows linearly with $T$ for fixed $W$, the probability of
anomaly is about $K\cdot e^{-E/(2N_0)}$, and so, the total MSE behaves
(see \cite[eq.\ (8.100), p.\ 633]{WJ65}) roughly according to
\begin{equation}
\label{totalmse}
\bE\{(\hat{m}-m)^2\}\approx\frac{N_0}{2W^2E}+B\cdot K
e^{-E/(2N_0)},
\end{equation}
where $B> 0$ is some constant, 
the first term accounts for the high--SNR MSE, and the second term is
the MSE dictated by the probability of an anomalous error. 
Note that here the degradation contributed by the anomalous error,
as a function of $N_0$, is graceful, in other words, there is still no sharp
breakdown of the kind that was described in the previous paragraph. This is
because of the fact that as long as $W$ 
is fixed, the $K=2WT$ orthonormal basis functions may capture only
a very small fraction of the `problematic 
directions' (as described in the previous
paragraph) of the entire plethora of `directions' of the noise, which is of
infinite bandwidth. In other words, since 
the probability of a large noise projection in a certain
direction is exponentially small, it takes exponentially many directions to
make the probability of a large projection in {\it at least} one of them,
considerably large.
As the energies
$E$ and $\dot{E}$, grow linearly with $T$ 
(for fixed power and bandwidth), the first term in (\ref{totalmse})
is proportional to $1/T$ while the second term decays exponentially in $T$. 
A natural question that arises then
is whether there may be better trade-offs. 
The answer is affirmative if $W$ would be
allowed to grow (exponentially fast) with $T$. Assuming then that
$W\propto e^{RT}$ for some fixed parameter $R > 0$, the first term would then
decay at the rate of $e^{-2RT}$ whereas the second term may still continue to
decay exponentially as long as $R$ is not too large. The exact behavior
depends, of course, on the form of the parametric family of signals
$\{s_m(t)\}$, but for some classes of signals like those pertaining to 
FPM, it is shown in \cite{WJ65} that the
probability of anomaly decays according to $e^{-TE(R)}$,
where $E(R)$ is the error exponent function 
pertaining to infinite--bandwidth orthogonal signals
over the additive white Gaussian noise (AWGN) channel, i.e.,
\begin{equation}
E(R)=\left\{\begin{array}{ll}
\frac{C}{2}-R & R < \frac{C}{4}\\
(\sqrt{C}-\sqrt{R})^2 & \frac{C}{4}\le R < C
\end{array}\right.
\end{equation}
Note that the best compromise between high-SNR MSE and anomalous MSE
pertains to 
the solution to the equation $E(R)=2R$, namely, $R=C/6$.
For $R > C$, the probability of anomaly tends to $1$ as $T\to \infty$. 
Thus, we observe that in the regime of unlimited 
bandwidth, the threshold effect pertaining to
anomalous errors is indeed sharp, while in the band--limited case, it is not.

Our purpose, in this work, is to study the threshold effect of anomalous
errors, in the unlimited bandwidth regime,
from a physical point of view, by relating the threshold effect to {\it phase
transitions} of large physical systems subjected to disorder, in particular,
a REM--like model, as described in the Introduction. The limit of
large $T$ would then correspond to the thermodynamic limit of a large system,
customarily considered in statistical physics.
Moreover, as discussed earlier, the physical point of view will help us to understand situations
where there is more than one phase transition.

\section{A Physical Perspective on the Threshold Effect}
\label{phys}

For the sake of concreteness, we consider the case where the parameter $m$ is
time delay, defined in units
of $T$.\footnote{More general situations will be discussed in the
sequel.}
Let then
\begin{equation}
r(t)=s(t-mT)+n(t),~~~-\frac{T}{2}\le t\le +\frac{T}{2},~~~-M\le
m\le+M,~~~M<\frac{1}{2}.
\end{equation}
We will also assume
that the signal
autocorrelation function, i.e.,
\begin{equation}
R_s(\tau)\dfn\int_{-T/2}^{+T/2}\mbox{d}ts(t)s(t+\tau),
\end{equation}
vanishes outside the interval $[-\Delta,+\Delta]$. In this case, it is natural
to define the anomalous error event as the event where the absolute value of
the estimation error,
$|\hat{m}-m|$, exceeds $\Delta$. Since the signal energy is $E$, then so is
$R_s(0)$. Assuming that the signal support lies entirely within the
interval $[-T/2,+T/2]$ for all allowable values of $m$ (i.e., $M\le 
\frac{1}{2}-\Delta/T$), the energy of $\{s(t-mT)\}$ is independent of $m$,
and then maximum likelihood estimation is equivalent to maximum correlation:
\begin{equation}
\hat{m}=\mbox{arg}\max_{m:~|m|\le M}
\int_{-T/2}^{+T/2}\mbox{d}t r(t)s(t-mT).
\end{equation}
If one treats $m$ as a uniformly distributed random variable, the
corresponding
posterior density of $m$ given $\{r(t),~-T/2\le t\le T/2\}$ is given by
\begin{eqnarray}
P(m|\{r(t),~-T/2\le t\le T/2\})&=&
\frac{\exp\left\{-\frac{1}{N_0}
\int_{-T/2}^{+T/2}[r(t)-s(t-mT)]^2\mbox{d}t\right\}}
{\int_{-M}^{+M}\mbox{d}m'\exp\left\{-\frac{1}{N_0}
\int_{-T/2}^{+T/2}[r(t)-s(t-m'T)]^2\mbox{d}t\right\}}\nonumber\\
&=&\frac{\exp\left\{\frac{2}{N_0}
\int_{-T/2}^{+T/2}r(t)s(t-mT)\mbox{d}t\right\}}
{\int_{-M}^{+M}\mbox{d}m'\exp
\left\{\frac{2}{N_0}\int_{-T/2}^{+T/2}r(t)s(t-m'T)\mbox{d}t\right\}}
\end{eqnarray}
where in the second equality, we have cancelled out the factor
$\exp\{-\frac{1}{N_0}\int_{-T/2}^{+T/2}r^2(t)\mbox{d}t\}$, which
appears both in the numerator and the denominator, and
we have used again the fact that the energy, $E$, of
$\{s(t-mT)\}$ is independent of $m$. 
Owing to the exponential form of this posterior distribution,
it can be thought of, 
in the language of statistical mechanics,
as the Boltzmann distribution with inverse 
temperature $\beta=2/N_0$ and Hamiltonian 
(i.e., energy as a function of $m$):
\begin{equation}
\calH(m)=-\int_{-T/2}^{+T/2}\mbox{d}t r(t)s(t-mT).
\end{equation}
This statistical--mechanical point of view suggests 
to expand the scope and define a family of probability distributions
parametrized by $\beta$, as follows:
\begin{equation}
P_\beta(m|\{r(t),~-T/2\le t\le T/2\})
=\frac{\exp\left\{\beta\int_{-T/2}^{+T/2}r(t)s(t-mT)\mbox{d}t\right\}}
{\int_{-M}^{+M}\mbox{d}m'
\exp\left\{\beta\int_{-T/2}^{+T/2}r(t)s(t-m'T)\mbox{d}t\right\}}
\end{equation}
There are at least three meaningful choices of 
the value of the parameter $\beta$:
The first is $\beta=0$, corresponding to the 
uniform distribution on $[-M,+M]$, which is the prior.
The second choice is $\beta=2/N_0$, 
which corresponds to the true posterior distribution, as said. Finally,
as $\beta\to\infty$, the density 
$P_\beta(\cdot| \{r(t),~-T/2\le t\le T/2\})$ 
puts more and more weight on the value of
$m$ that maximizes the correlation $\int_{-T/2}^{+T/2}\mbox{d}t r(t)s(t-mT)$,
namely, on the ML estimator $\hat{m}$. It should be emphasized that if we
vary the parameter $\beta$, this is not 
necessarily equivalent to a corresponding
variation in the choice of $N_0$, according to $\beta=2/N_0$. 
For example, one may
examine the behavior of the ML estimator by letting $\beta\to\infty$, but
still analyze its performance for a given finite value of $N_0$.
This is to say that $P_\beta(\cdot| \{r(t),~-T/2\le t\le T/2\})$ should
only be thought of as an {\it auxiliary} 
posterior density function, not as the real one.
The denominator
of $P_\beta(m|\{r(t),~-T/2\le t\le T/2\})$, namely,
\begin{equation}
\zeta(\beta)=\int_{-M}^{+M}\mbox{d}m
\exp\left\{\beta\int_{-T/2}^{+T/2}r(t)s(t-mT)\mbox{d}t\right\}
\end{equation}
can then be thought of as the partition function pertaining to the
Boltzmann distribution 
$P_\beta(\cdot| \{r(t),~-T/2\le t\le T/2\})$.

Now, without essential loss of generality, let us assume that 
the true parameter value is $m=0$, that $\Delta$ divides $2MT$,
and that the integer $K=2MT/\Delta$ is an even number.
Consider the partition of the interval $[-M,+M]$ of possible values of
$m$ into sub-intervals of size $\Delta/T$. Let
$\calM_i=[i\Delta/T,(i+1)\Delta/T)$ denote the $i$--th sub-interval,
$i=-K/2,-K/2+1,\ldots-1,0,+1,\ldots,~K/2$. We will find it convenient to view 
the ML estimation of $m$ as a two--step procedure, where one first maximizes
the correlation $\int_{-T/2}^{+T/2}\mbox{d}t r(t)s(t-mT)$ within each
sub-interval $\calM_i$, i.e., calculate
\begin{equation}
\max_{m\in\calM_i}\int_{-T/2}^{+T/2}\mbox{d}t r(t)s(t-mT),
\end{equation}
and then take the largest maximum over all $i$.
Let us define
\begin{equation}
\epsilon_0=\max_{|m|\le \Delta/T}\int_{-T/2}^{+T/2}\mbox{d}t r(t)s(t-mT)=
\max_{m\in\calM_0\cup\calM_{-1}}\int_{-T/2}^{+T/2}\mbox{d}t
r(t)s(t-mT)
\end{equation}
and for $i\ne 0$,
\begin{equation}
\epsilon_i=\max_{m\in\calM_i}\int_{-T/2}^{+T/2}\mbox{d}t
r(t)s(t-mT),~~~~1\le i\le K/2-1
\end{equation}
\begin{equation}
\epsilon_i=\max_{m\in\calM_{i-1}}\int_{-T/2}^{+T/2}\mbox{d}t
r(t)s(t-mT),~~~~-(K/2-1)\le i \le -1
\end{equation}
Thus, for the purpose of analyzing the behavior of the ML
estimator, we can use a modified version of the partition function,
defined as
\begin{equation}
Z(\beta)=\sum_{i=-K/2+1}^{K/2-1} e^{\beta\epsilon_i},
\end{equation}
and analyze it in the limit of $\beta\to\infty$ (the low temperature limit).
Note that here, $\epsilon_i$ has the 
meaning of the (negative) Hamiltonian pertaining to
a `system configuration' indexed by $i$.

In order to characterize the behavior of $Z(\beta)$, it is instructive
to recognize that it is quite similar to the random energy model
(REM) of disordered spin glasses: According to the REM, the energies
$\{\epsilon_i\}$, pertaining
to various system configurations indexed by $i$, are i.i.d.\ random variables,
normally assumed zero--mean and Gaussian, 
but other distributions are possible too. This is not quite exactly our case,
but as we shall see shortly, this is close enough to allow the techniques
associated with the analysis of the REM to be applicable here.

First, observe that under these assumptions,
\begin{eqnarray}
\epsilon_0&=&\max_{|m|\le \Delta/T}\int_{-T/2}^{+T/2}\mbox{d}t
[s(t)+n(t)]s(t-mT)\nonumber\\
&=&\max_{|m|\le
\Delta/T}\left[R_s(mT)+\int_{-T/2}^{+T/2}\mbox{d}tn(t)s(t-mT)\right]
\end{eqnarray}
whereas for $i\ne 0$,
\begin{equation}
\epsilon_i=\max_{m\in\calM_i}\int_{-T/2}^{+T/2}
\mbox{d}tn(t)s(t-mT),~~~~~i>0,
\end{equation}
and
\begin{equation}
\epsilon_i=\max_{m\in\calM_{i-1}}
\int_{-T/2}^{+T/2}\mbox{d}tn(t)s(t-mT),~~~~~i<0.
\end{equation}
As for $\epsilon_0$, we have, on the one hand
\begin{equation}
\epsilon_0 \ge
R_s(0)+\int_{-T/2}^{+T/2}\mbox{d}tn(t)s(t)=
PT+\int_{-T/2}^{+T/2}\mbox{d}tn(t)s(t)
\end{equation}
and on the other hand,
\begin{equation}
\epsilon_0\le \max_{|m|\le\Delta/T} R_s(mT)+
\max_{|m|\le\Delta/T} \int_{-T/2}^{+T/2}\mbox{d}tn(t)s(t-mT)=
PT+\max_{|m|\le\Delta/T} \int_{-T/2}^{+T/2}\mbox{d}tn(t)s(t-mT).
\end{equation}
Considering the limit $T\to\infty$ for fixed $P$, 
both the upper bound and the lower bound are dominated by the first term,
which grows linearly with $T$, while the second term is a random variable
whose standard deviation, for large $T$, scales 
in proportion to $\sqrt{T}$.
Thus, for a typical realization of $\{n(t), -T/2\le t\le T/2\}$,
$\epsilon_0\approx PT$, and so, its typical contribution to the partition
function is given by
\begin{equation}
Z_0(\beta)\dfn e^{\beta\epsilon_0}\approx e^{\beta PT}.
\end{equation}
Consider now the contribution of all the other $\{\epsilon_i\}$ to 
the partition function, and define
\begin{equation}
Z_a(\beta)=\sum_{i\ne 0}e^{\beta\epsilon_i},
\end{equation}
where the subscript $a$ stands for `anomaly', as this term pertains to
anomalous errors. The total partition function is, of course,
\begin{equation}
Z(\beta)=Z_0(\beta)+Z_a(\beta).
\end{equation}
Now, for $i\ne 0$, $\{\epsilon_i\}$ are identically distributed RV's, which
are alternately independent, i.e.,
$\ldots,\epsilon_{-3},\epsilon_{-1},\epsilon_1,\epsilon_3,\ldots$ are
independent (since the noise is white and $R_s(\tau)$ vanishes
for $|\tau|\ge \Delta$),
and so are $\ldots,\epsilon_{-4},\epsilon_{-2},\epsilon_2,\epsilon_4,\ldots$.
In order to evaluate the typical behavior of $Z_a(\beta)$,
we shall represent it as
\begin{equation}
Z_a(\beta)=\int\mbox{d}\epsilon N(\epsilon)e^{\beta\epsilon},
\end{equation}
where $N(\epsilon)\mbox{d}\epsilon$ is the number of $\{\epsilon_i\}$ that
fall between $\epsilon$ and $\epsilon+\mbox{d}\epsilon$, i.e.,
\begin{equation}
N(\epsilon)\mbox{d}\epsilon=\sum_{i\ne
0}\calI(\epsilon\le\epsilon_i\le\epsilon+\mbox{d}\epsilon),
\end{equation}
where $\calI(\cdot)$ is the indicator function of an event.
Obviously, 
\begin{equation}
\bE\{N(\epsilon)\mbox{d}\epsilon\}=
K\cdot\mbox{Pr}\{\epsilon\le\epsilon_i\le\epsilon+\mbox{d}\epsilon\}
\end{equation}
and so,
\begin{equation}
\bE\{N(\epsilon)\}=K\cdot f(\epsilon),
\end{equation}
where $f(\epsilon)$ is the probability density function (pdf) of $\epsilon_i$,
for $i\ne 0$.
Now, to accommodate the asymptotic
regime of $W\propto e^{RT}$, we take the signal duration to be
$\Delta=\Delta_0e^{-RT}$, where $\Delta_0>0$ is a fixed parameter, and so,
\begin{equation}
K=\frac{2MT}{\Delta_0}\cdot e^{RT}.
\end{equation}
Thus, $N(\epsilon)\mbox{d}\epsilon$ is the 
sum of exponentially many binary random
variables. As said earlier, although these random variables are not
independent, they are alternately independent, and so, if
$N(\epsilon)\mbox{d}\epsilon$ is represented as
\begin{equation}
\sum_{i\ne 0~\mbox{even}}\calI(\epsilon\le
\epsilon_i\le\epsilon+\mbox{d}\epsilon)+
\sum_{i~\mbox{odd}}\calI(\epsilon\le\epsilon_i\le\epsilon+\mbox{d}\epsilon)
\end{equation}
then each of the two terms is the sum of i.i.d.\ binary random variables,
whose typical value is zero when 
$\frac{K}{2}\cdot f(\epsilon)\mbox{d}\epsilon << 1$
and $\bE\{N(\epsilon)\mbox{d}\epsilon\}$ when $\frac{K}{2}\cdot
f(\epsilon)\mbox{d}\epsilon >> 1$. This means that, asymptotically, for large
$T$, only energy levels for
which $\ln f(\epsilon) > -RT$ will typically be populated by some $\{i\}$.
Let $\varepsilon_T$ be the largest solution to 
the equation $\ln f(\epsilon) >
-RT$. 
Then, the typical value of $Z_a$ is exponentially
\begin{eqnarray}
Z_a(\beta,R)&\exe&\int_{-\infty}^{\varepsilon_T}\frac{T}{\Delta_0}\cdot
e^{RT}f(\epsilon)e^{\beta\epsilon}\mbox{d}\epsilon\nonumber\\
&\exe&\exp\left\{\max_{\epsilon\le\varepsilon_T}\left[RT+\ln
f(\epsilon)+\beta\epsilon\right]\right\}
\end{eqnarray}
where $\exe$ denotes asymptotic equality in the exponential scale\footnote{For
two non--negative functions $a(T)$ and $b(T)$, the notation $a(T)\exe b(T)$
means that $\lim_{T\to\infty}\frac{1}{T}\ln\frac{a(T)}{b(T)}=0$.}
as $T\to\infty$, and
where we have modified the notation from $Z_a(\beta)$ to $Z_a(\beta,R)$
to emphasize the dependence on the exponential growth rate $R$ of the
parameter $K$.
Any further derivation, from this point onward, requires the knowledge 
of the pdf $f(\epsilon)$, which is known accurately only for
certain specific choices of the pulse shape. One of them, that we will assume
here for concreteness, is the rectangular pulse
\begin{equation}
s(t)=\left\{\begin{array}{ll}
\sqrt{\frac{E}{\Delta}} & |t|\le \frac{\Delta}{2}\\
0 & \mbox{elsewhere}\end{array}\right.
\end{equation}
where $E=PT$, $P$ being the average power of the signal.
Therefore,
\begin{equation}
R_s(\tau)=E\left[1-\frac{|\tau|}{\Delta}\right]_+=
PT\left[1-\frac{|\tau|}{\Delta}\right]_+
\end{equation}
where $[x]_+\dfn\max\{0,x\}$.
From a result by Slepian \cite{Slepian62} 
in a form that was later derived by Shepp
\cite{Shepp66}
(see also \cite{ZZ69}),
it is known that if $X_\theta$ is a zero--mean Gaussian random process with
autocorrelation function $R(\tau)=[1-|\tau|]_+$, then the cumulative probability
distribution function of $Y=\sup_{0\le \theta\le 1}X_\theta$ is given by
\begin{equation}
F_0(a)=\mbox{Pr}\{Y\le
a\}=[1-\Phi(a)]^2-\frac{ae^{-a^2/2}}
{\sqrt{2\pi}}[1-\Phi(a)]-\frac{e^{-a^2}}{2\pi}
\end{equation}
where
\begin{equation}
\Phi(a)\dfn\frac{1}{\sqrt{2\pi}}\int_a^\infty e^{-u^2/2}\mbox{d}u.
\end{equation}
This means that the density of $Y$ is given by
\begin{equation}
f_0(a)=\frac{\mbox{d}F_0(a)}{\mbox{d}a}=
\frac{ae^{-a^2}}{2\pi}+[1-\Phi(a)](1+a^2)\frac{e^{-a^2/2}}{\sqrt{2\pi}}.
\end{equation}
This result applies, in our case, to the random process
\begin{equation}
X_\theta=\sqrt{\frac{2}{N_0PT}}\cdot\int_{-T/2}^{+T/2}
\mbox{d}tn(t)s(t-\theta\Delta),~~~~0\le\theta\le 1,
\end{equation}
which means that for $i\ne 0$, the probability density
function of $\epsilon_i$ is given by
\begin{equation}
f(\epsilon)=\sqrt{\frac{2}{N_0PT}}\cdot
f_0\left(\frac{\epsilon}{\sqrt{N_0PT/2}}\right).
\end{equation}
Thus,
\begin{equation}
Z_a(\beta,R)\exe\exp\left\{RT+\frac{1}{2}\ln\left(\frac{2}{N_0PT}\right)+
\max_{\epsilon\le\varepsilon_T}\left[\ln
f_0\left(\frac{\epsilon}{\sqrt{N_0PT/2}}\right)+\beta\epsilon\right]\right\}.
\end{equation}
Now, the exact form of $f_0$ may not lend itself to convenient analysis, but
considering the asymptotic limit of $T\to\infty$, it is not difficult to see
(due to the scaling by $\sqrt{N_0PT/2}$ in the argument of $f_0(\cdot)$)
that the maximum at the exponent of the last expression is attained for values
of $\epsilon$ that grow without bound as $T\to\infty$. It would therefore be
convenient to approximate $f_0(a)$ given above by its dominant term for very
large $a$, which is given by
\begin{equation}
f_0(a)\approx \frac{a^2e^{-a^2/2}}{\sqrt{2\pi}}.
\end{equation}
On substituting this approximation, 
we first find an approximation to $\varepsilon_T$ according to
\begin{equation}
2\ln\left(\frac{\varepsilon_T}{\sqrt{N_0PT}}\right)-
\frac{\varepsilon_T^2}{N_0PT}=-RT.
\end{equation}
For large $T$, the first term is negligible compared to the second term and
the right--hand side, and so, $\varepsilon_T$ is well approximated as
\begin{equation}
\varepsilon_T=\sqrt{N_0PR}\cdot T.
\end{equation}
Next, we use the approximate form of $f_0$ in the maximization of $\ln
f_0(\epsilon/\sqrt{N_0PT/2})+\beta\epsilon$, i.e., solve the problem
\begin{equation}
\max_{\epsilon\le\sqrt{N_0PR}T}
\left[\ln\left(\frac{2\epsilon^2}{N_0PT}\right)-\frac{\epsilon^2}
{N_0PT}+\beta\epsilon\right]
\end{equation}
whose maximizer, for large $T$, is easily found to be approximated by
\begin{equation}
\epsilon_*=\min\left\{\sqrt{N_0PR}\cdot T,\frac{\beta N_0PT}{2}\right\}.
\end{equation}
On substituting this back into the expression of $Z_a(\beta,R)$, 
and defining 
\begin{equation}
\psi_a(\beta,R)=\lim_{T\to\infty}\frac{\ln Z_a(\beta,R)}{T}, 
\end{equation}
we get
\begin{equation}
\psi_a(\beta,R)=\left\{\begin{array}{ll}
R+\frac{\beta^2N_0P}{4} & \beta <
\beta_c(R)\\
\beta\sqrt{N_0PR} & \beta >
\beta_c(R)\end{array}\right.
\end{equation}
where
\begin{equation}
\beta_c(R)=\frac{2}{N_0}\sqrt{\frac{R}{C}},
\end{equation}
$C=P/N_0$ being the capacity of infinite--bandwidth AWGN channel.
Thus, we see that $Z_a(\beta,R)$ undergoes a phase transition at
$\beta=\beta_c(R)$: For
$\beta < \beta_c(R)$, 
$Z_a(\beta,R)$ is dominated by an exponential number of
$\{i\}$ for which $\epsilon_i$ is about $\beta N_0 PT/2$. 
As $\beta$ exceeds $\beta_c(R)$, the system pertaining to $Z_a$ undergoes a
phase transition,
where $Z_a(\beta,R)$ becomes dominated by a sub-exponential number
of $\{i\}$ at the `ground state' level of $\sqrt{N_0PR}\cdot T$.
This sub-exponential number of dominant ground--state `configurations'
corresponds to a zero entropy, yet disordered phase, which is called
in the terminology of physicists, the {\it glassy phase}
(see \cite[Chap.\ 5]{MM09}).

\begin{figure}[ht]
\hspace*{3cm}\input{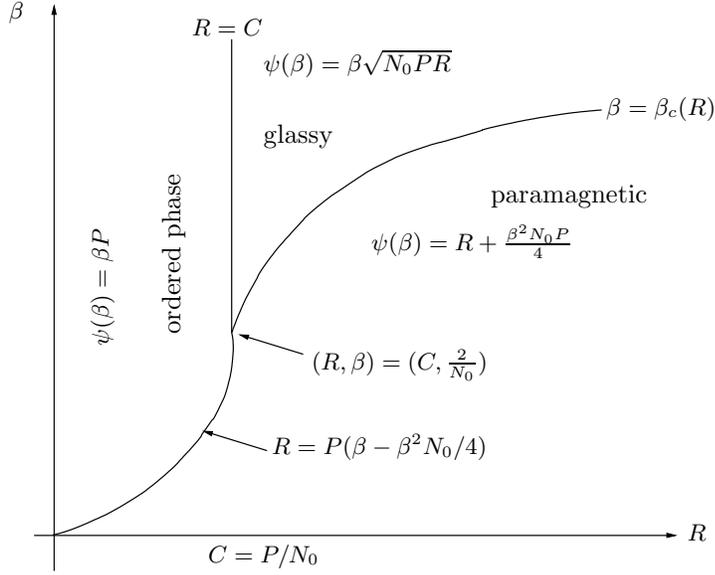}
\caption{\small Phase diagram for ML estimation of a delay parameter.}
\label{fig1}
\end{figure}

Taking now into account the contribution of $Z_0(\beta)$, 
and defining 
\begin{equation}
\psi(\beta,R)=\lim_{T\to\infty}\frac{\ln Z(\beta,R)}{T},
\end{equation}
we end up with three
phases, as can be seen in the following expression
\begin{equation}
\psi(\beta,R)=\left\{\begin{array}{ll}
\beta P & \{R <P(\beta-\beta^2N_0/4),~\beta<2/N_0\}\bigcup
\{R< C,~\beta\ge 2/N_0\}\\
R+\frac{\beta^2N_0P}{4} & \{R> C,~\beta
<\beta_c(R)\}\bigcup\{P(\beta-\beta^2N_0/4)<R<C,~\beta<2/N_0\}\\
\beta\sqrt{N_0PR} & \mbox{elsewhere}
\end{array}\right.
\end{equation}
The phase diagram is depicted in Fig.\ \ref{fig1}.
As said earlier, for ML estimation, the relevant regime is $\beta\to\infty$,
where as can be seen, the system undergoes a phase transition as $R$ exceeds
$C$. This phase transition captures the threshold effect in the estimation of
the delay parameter $m$, in this example.

As long as $R < C$, the probability of anomaly is still vanishingly small, and
the dominant event is that of a small error (less than $\Delta/T$ in absolute
value). The critical point, where all three phases meet, is the point
$(C,2/N_0)$. Note that $\beta=2/N_0$ is the `natural' value of $\beta$ that
arises in the true posterior of $m$ given $\{r(t),~-T/2\le t\le T/2\}$.

As we can see, the physical perspective provides some insight, not only
concerning the estimation of the parameter $m$, but moreover, about the
posterior of $m$ given the noisy signal $\{r(t),~-T/2\le t\le T/2\}$.
If we use the `correct' value of $\beta$ or larger
i.e., $\beta\ge 2/N_0$,
then as long as $R < C$,
the posterior possesses a very sharp peak around the true value of $m$
and the width of this peak does not exceed $\Delta/T$ from either side.
This is the {\it ordered phase}, or the {\it 
ferromagnetic phase}, in the jargon
of physicists.
As $R$ crosses $C$, then the behavior is as follows: If $\beta=2/N_0$,
the posterior changes abruptly and instead of one peak around the true $m$,
it becomes dominated by exponentially many `spikes' scattered across the
whole interval $[-M,+M]$. This is the {\it paramagnetic phase}.
If, on the other hand, $\beta > 2/N_0$, then
there is an intermediate range of rates $R\in[C,\beta^2N_0P/4]$, where
the number of such spikes 
is still sub--exponential, which means the glassy phase.
Finally, as one continues to increase $R$ above $\beta^2N_0P/4$, the number of
spikes becomes exponential (the paramagnetic phase). 

On the other hand, for $\beta
< 2/N_0$, the abrupt transition to exponentially many spikes happens for $R
=P(\beta-\beta^2N_0/4)$, which is less than $C$. The fixed bandwidth regime
corresponds to the vertical axis ($R=0$) 
in the phase diagram, and as can be
seen, no phase
transition occurs along this axis at any finite temperature.
This is in agreement with our earlier discussion on the graceful behavior of
the probability of anomaly
at fixed bandwidth.

It is instructive to compare the behavior of the ML estimator to
the Weiss--Weinstein lower bound \cite{WW85}, \cite{Weiss85}
because this bound is claimed to
capture the threshold effect. As we have seen, the ML estimator
has the following ranges of exponential behavior as a function of $R$:
\begin{equation}
\bE\{(\hat{m}-m)^2\} \sim \left\{\begin{array}{ll}
e^{-2RT} & R < C/6\\
e^{-E(R)T} & C/6 < R < C\\
e^{-0\cdot T} & R > C \end{array}\right.
\end{equation}
On the other hand, the
Weiss--Weinstein bound (WWB) for estimating a rectangular
pulse in Gaussian white noise is given (in our notation) by
\begin{equation}
\mbox{WWB}=\max_{h\ge 0}\frac{h^2[1-h/T]_+^2
\exp\{-[h/\Delta]^+CT/2\}}
{2\left[1-(1-2h/T)_+\exp\{-[h/\Delta]^+CT/2\}\right]},
\end{equation}
where $[x]_+=\max\{x,0\}$ and
$[x]^+=\min\{x,1\}$.
Examining this bound under the asymptotic regime of $T\to\infty$ with
$\Delta=\Delta_0e^{-RT}$, yields the following behavior:
\begin{equation}
\mbox{WWB}\sim\left\{\begin{array}{ll}
e^{-2RT} & R < C/4\\
e^{-CT/2} & R > C/4 \end{array}\right.
\end{equation}
In agreement with the analysis in \cite{Weiss85},
we readily observe that for a given
$R$ and
for high SNR ($C=P/N_0\to\infty$), both quantities are of the exponential
order of
$e^{-2RT}$, whereas
for low SNR ($C\to 0$), both are about $e^{-CT/2}\sim e^{-0\cdot T}$. However,
if we look at both quantities as functions of $R$ for fixed
$C > 0$, there is a different behavior. Not only the phase
transition points differ, but also the large $R$
asymptotics disagree. Thus, the WWB
indeed captures the threshold effect of
the ML estimator, but in a slightly
weaker sense when it comes to the asymptotic wide-band regime.

\subsubsection*{Discussing Some Extensions}

It is interesting to slightly expand the scope 
to a situation of mismatched estimation.
Suppose that instead of ML estimation based on the known waveform $s(t)$,
the estimator is based on maximizing the temporal correlation with another
waveform, $\tilde{s}(t-mT)$, whose energy is $E=PT$ and whose width is
$\tilde{\Delta}=\Delta_0e^{-\tilde{R}T}$. In this case, the phase diagram,
in the plane of $\beta$ vs.\ $\tilde{R}$, 
will remain essentially the same as in Fig.\ \ref{fig1}, except that
there will be a degradation by a factor of $\rho$ in $\beta$, and by a factor 
of $\rho^2$ in the rate, where
\begin{equation}
\rho\dfn\frac{1}{E}\int_{-T/2}^{+T/2}s(t)\tilde{s}(t)\mbox{d}t.
\end{equation}
In other words, the triple point will be $(\rho^2C,2\rho/N_0)$, the
vertical straight--line ferromagnetic--glassy 
phase boundary will be $\tilde{R}=\rho^2C$,
rather than $R=C$. The other
phase boundaries will be as follows: the paramagnetic--ferromagnetic 
boundary is the parabola
$\tilde{R}=P(\rho\beta-\beta^2N_0/4)$, and the paramagnetic--glassy boundary
would continue to be the parabola $\beta=\beta_c(\tilde{R})$, where
the function $\beta_c(\cdot)$ is as defined before. The dependence on the
parameter $R$ of the real signal is solely via its effect on the parameter
$\rho$.

Our derivations above are somewhat specific to the example of time delay
estimation, and for the special case of a rectangular pulse. Therefore, a
few words about the more general picture are in order. 
First, consider time delay estimation of more general
signals. We assumed that
$R_s(\tau)$ vanishes for $|\tau|\ge\Delta$, but this still leaves room for
more general pulses with support $\Delta$, not necessarily the rectangular
one. Unfortunately, as said earlier, the exact pdf of $\epsilon_i$, $i\ne 0$,
is not known for a general autocorrelation function that is induced by 
a general
choice of $s(t)$. However, for our asymptotic analysis in the regime of
$T\to\infty$, what counts (as we have seen) is actually merely the tail 
behavior of this pdf, and this tail is known, under fairly general conditions
(see \cite[p.\ 40]{Adler90}, with
a reference also to \cite{MP71}),
to behave the same way as the tail of the Gaussian pdf of
zero mean and variance $N_0PT/2$. Therefore, our approximate analysis
in the large $T$ limit would continue to apply for other pulse shapes
as well.

Second, consider the estimation of parameters other than delay (e.g., frequency
offset or phase), still requiring that the time correlation between
$s_m(t)$ and $s_{m'}(t)$ would essentially vanish whenever $|m-m'|$ exceeds
a certain threshold (in our earlier example, $\Delta/T$).
In this case, as we have seen, the high--SNR MSE is inversely
proportional to the squared norm, $\dot{E}$,
of the vector $\dot{\bs}(m)$ of derivatives of
$\{s_i(m)\}$ w.r.t.\ $m$. Again, 
assuming that this norm is independent of $m$, it is
proportional to the length of the signal locus, as discussed earlier. 
For a good trade-off between the high--SNR MSE and the anomalous MSE, we would
like to modulate the parameter in such a way that for a given $E$, the
quantity $\dot{E}$ would grow exponentially with $T$, i.e.,
$\dot{E}\propto e^{2RT}$, as an extension of our earlier discussion in the
case of a time delay. For example, in the case of frequency--position
modulation, where $s(t)=A\cos(2\pi(f_c+mW)t+\phi)$, $|m|\le M$, 
$W<<f_c$, both $f_c$ and $W$ should be proportional to $e^{RT}$.
The corresponding analysis of $\epsilon_i$ and the associated
partition function would be, in principle, 
similarly as before, except that one should
consider the process 
$X_\theta=\int_{-T/2}^{+T/2}n(t)\cos(2\pi(f_c+\theta W)t+\phi)
\mbox{d}t$, and the remarks of the previous paragraph continue to apply.
Similar comments apply to other kinds of parametrization.

\section{Joint ML Estimation of Amplitude and Delay}
\label{joint}

\begin{figure}[ht]
\hspace*{3cm}\input{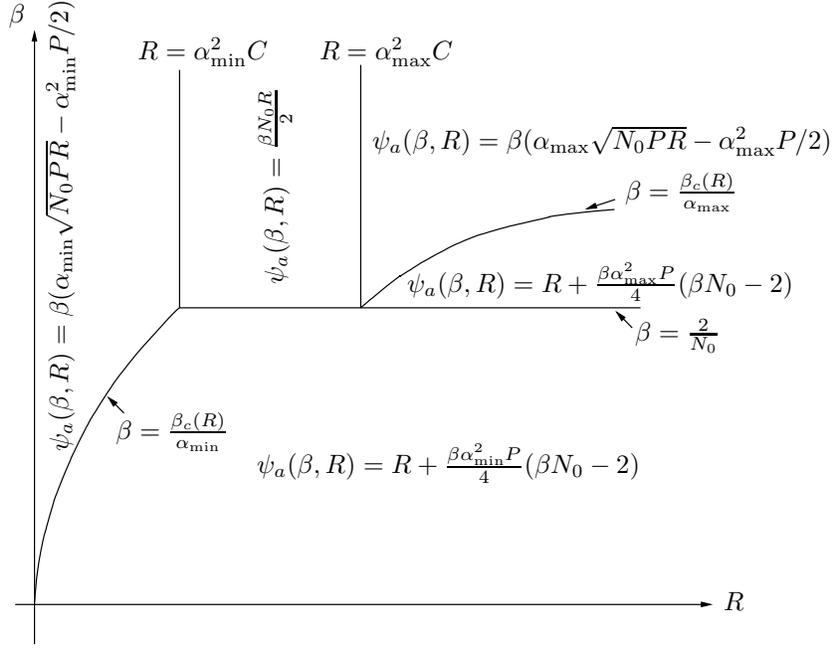}
\caption{\small Phase diagram of $Z_a(\beta,R)$ for joint ML estimation
of amplitude and delay.}
\label{fig1a}
\end{figure}

We now extend our earlier study to the model
\begin{equation}
r(t)=\alpha\cdot s(t-mT)+n(t),~~~~-\frac{T}{2}\le t\le +\frac{T}{2}
\end{equation}
where now both $\alpha$ and $m$ are parameters to be estimated, and where it is
assumed that $m\in[-M,+M]$ 
as before and $\alpha\in[\alpha_{\min},\alpha_{\max}]$,
with $0< \alpha_{\min}\le 1\le \alpha_{\max}$ and
\begin{equation}
\frac{1}{\alpha_{\max}-\alpha_{\min}}\cdot
\int_{\alpha_{\min}}^{\alpha_{\max}}\alpha^2\mbox{d}\alpha=1 
\end{equation}
which means that
the average energy (w.r.t.\ the uniform distribution within
the interval $[\alpha_{\min},\alpha_{\max}]$)
of the received signal is still $E$.
Here the energy of the received signal depends on $\alpha$, as it is given by
$\alpha^2E$. The relevant partition 
function would be
\begin{eqnarray}
Z(\beta,R)=
\int_{\alpha_{\min}}^{\alpha_{\max}}\mbox{d}\alpha\sum_i
\exp[\beta(\alpha\epsilon_i-\alpha^2PT/2)]
\dfn \int_{\alpha_{\min}}^{\alpha_{\max}}\mbox{d}\alpha Z(\alpha,\beta,R).
\end{eqnarray}
The analysis of $Z_a(\alpha,\beta,R)$ (which is 
the same expression except that the
sum excludes $i=0$)
in the framework of a REM--like model, is precisely the same as before
except that $\beta$ is replaced by $\beta\alpha$ and there is another
multiplicative factor of
$\exp\{-\beta \alpha^2PT/2\}$.
Accordingly, re--defining 
\begin{equation}
\psi_a(\alpha,\beta,R)=\lim_{T\to\infty}\frac{\ln
Z_a(\alpha,\beta,R)}{T},
\end{equation}
we get the following results: For $\beta
\le\beta_c(R)/\alpha_{\max}$,
\begin{equation}
\psi_a(\alpha,\beta,R)=R+\frac{\beta\alpha^2P}{4}(\beta N_0-2)~~~~~~~\forall
\alpha_{\min}\le\alpha\le\alpha_{\max}.
\end{equation}
Similarly, $\beta \ge \beta_c(R)/\alpha_{\min}$
\begin{equation}
\psi_a(\alpha,\beta,R)=
\beta\left(\alpha\sqrt{N_0PR}-\frac{\alpha^2P}{2}\right)~~~~~~\forall
\alpha_{\min}\le\alpha\le\alpha_{\max}.
\end{equation}
Finally, for $\beta\in(\beta_c(R)/\alpha_{\max},\beta_c(R)/\alpha_{\min})$
we have:
\begin{equation}
\psi_a(\alpha,\beta,R)=\left\{\begin{array}{ll}
R+\frac{\beta\alpha^2P}{4}(\beta N_0-2) & \alpha_{\min}\le\alpha
\le\frac{2}{\beta N_0}\sqrt{\frac{R}{C}}\\
\beta\left(\alpha\sqrt{N_0PR}-\frac{\alpha^2P}{2}\right) &
\frac{2}{\beta
N_0}\sqrt{\frac{R}{C}}\le\alpha\le\alpha_{\max}\end{array}\right.
\end{equation}
Upon maximizing over $\alpha$, we get five different phases of
$\psi_a(\beta,R)=\max_\alpha\psi_a(\alpha,\beta,R)$, three glassy phases
and two paramagnetic ones:
\begin{equation}
\psi_a(\beta,R)=\left\{\begin{array}{ll}
\beta\left(\alpha_{\min}\sqrt{N_0PR}-\frac{\alpha_{\min}^2P}{2}\right) &
R < \alpha_{\min}^2C~\mbox{and}~\beta>\frac{\beta_c(R)}{\alpha_{\min}}\\
\frac{\beta N_0R}{2} & R\in(\alpha_{\min}^2C,\alpha_{\max}^2C)~\mbox{and}~
\beta>\frac{2}{N_0}\\
\beta\left(\alpha_{\max}\sqrt{N_0PR}-\frac{\alpha_{\max}^2P}{2}\right) &
R >\alpha_{\max}^2 C~\mbox{and}~\beta > \frac{\beta_c(R)}{\alpha_{\max}}\\
R+\frac{\beta\alpha_{\min}^2P}{4}(\beta N_0-2) & 
\beta\le\min\left\{\frac{\beta_c(R)}{\alpha_{\min}},\frac{2}{N_0}\right\}\\
R+\frac{\beta\alpha_{\max}^2P}{4}(\beta N_0-2) & R >
\alpha_{\max}^2C~\mbox{and}~
\beta\in\left(\frac{2}{N_0},\frac{\beta_c(R)}
{\alpha_{\max}}\right)\end{array}\right.
\end{equation}
In Figure \ref{fig1a}, we show the phase diagram of $\psi_a(\beta,R)$.
As can be seen, the paramagnetic phase is split into the two sub--phases,
according to $\beta < 2/N_0$ and $\beta > 2/N_0$, whereas the glassy phase
is split into three parts, according to the range of $R$.

Finally, when we take into account the contribution of $Z_0(\beta)=e^{\beta
PT/2}$, where
it is assumed that that true values of the parameters are $\alpha_0=1$ and
$m_0=0$,
we end up with the following expression for the re--defined
\begin{equation}
\psi(\beta,R)\dfn\lim_{T\to\infty}\frac{\ln Z(\beta,R)}{T}
\end{equation}
which is given by
\begin{equation}
\psi(\beta,R)=\left\{\begin{array}{ll}
\frac{\beta P}{2} & \left\{R < C~\mbox{and}~
\beta>\frac{2}{N_0}\right\}\bigcup\left\{
R<R_\beta~\mbox{and}~
\beta\le\frac{2}{N_0}\right\}\\
\frac{\beta N_0R}{2} & R\in(C,\alpha_{\max}^2C)~\mbox{and}~
\beta>\frac{2}{N_0}\\
\beta\left(\alpha_{\max}\sqrt{N_0PR}-\frac{\alpha_{\max}^2P}{2}\right) &
R >\alpha_{\max}^2 C~\mbox{and}~\beta > \frac{\beta_c(R)}{\alpha_{\max}}\\
R+\frac{\beta\alpha_{\max}^2P}{4}(\beta N_0-2) & R > \alpha_{\max}^2 C~
\mbox{and}~
\beta\in\left(\frac{2}{N_0},\frac{\beta_c(R)}{\alpha_{\max}}\right)\\
R+\frac{\beta\alpha_{\min}^2P}{4}(\beta N_0-2) & R > R_\beta~\mbox{and}~
\beta\le\frac{2}{N_0}
\end{array}\right.
\end{equation}
where
\begin{equation}
R_\beta\dfn \frac{P}{2}\left[\beta(1+\alpha_{\min}^2)-
\frac{\beta^2N_0\alpha_{\min}^2}{2}\right].
\end{equation}
The phase diagram of this function is depicted in Fig.\ \ref{fig1b}. 

\subsubsection*{Discussion}

Although the model is linear
in the parameter $\alpha$, its interaction with $m$
exhibits, in general, more phases than the parameter $m$ alone, and it
causes anomalies
in the estimation of $\alpha$ as well, but these anomalies have
a different character than those associated with $m$: While the 
anomaly makes the estimator of $m$ become an essentially uniformly
distributed random variable within the interval $[-M, +M]$, 
the anomalous estimator of $\alpha$ tends to concentrate on a
deterministic value as $T\to\infty$. To see why this is true, observe
that in the limit of large $\beta$ (which is
relevant for ML estimation), as long as $R < C$, the estimation
error is typically not anomalous.
For $C < R < \alpha_{\max}^2C$, the dominant value of
$\hat{\alpha}$ is $\sqrt{R/C}$, whereas for
$R > \alpha_{\max}^2C$, the dominant value of
$\hat{\alpha}$ is $\alpha_{\max}$. For low $\beta$, 
we also identify the region where the posterior of $(\alpha,m)$ is
dominated by points where $\alpha=\alpha_{\min}$.

\begin{figure}[ht]
\hspace*{3cm}\input{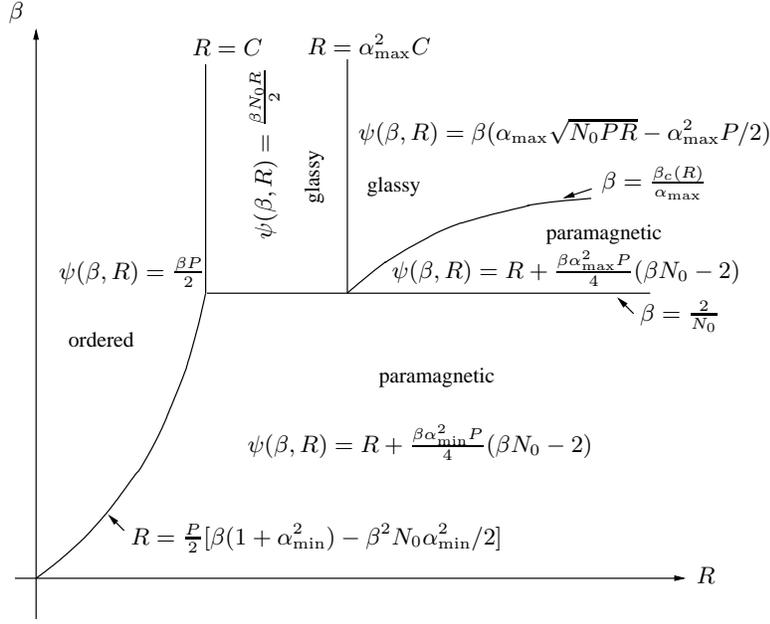}
\caption{\small Phase diagram of $Z(\beta,R)$ for joint ML estimation
of amplitude and delay.}
\label{fig1b}
\end{figure}

Referring to Fig.\ \ref{fig1b}, in 
the special case where $\alpha_{\max}=\infty$, 
the eastern glassy phase
and the northern paramagnetic phase disappear, and
we end up with three phases only: the ordered phase (unaltered), the southern
paramagnetic phase, and the western glassy phase.
If, in addition, $\alpha_{\min}=0$ 
(i.e., we know nothing a--priori on $\alpha$), then the curve
$R=R_\beta$ becomes a straight line ($R=\beta P/2$) and in the paramagnetic
region, we get $\psi(\beta,R)=R$. On the other hand, the case
$\alpha_{\min}=\alpha_{\max}=1$ (i.e., $\alpha=1$ and there is no uncertainty
in $\alpha$), we are back to the earlier case of a delay parameter only.

\section{Summary and Conclusion}
\label{conc}

In this paper, we proposed a statistical--mechanical perspective on
the threshold effect in parameter estimation of 
non--linearly modulated wide-band signals corrupted
by additive white Gaussian noise. The proposed framework, 
which is mapped into a
REM--like model of disordered spin glasses, provides a fairly comprehensive
picture of the behavior of the ML estimator as a function of the bandwidth
parameter $R$ and the temperature parameter $\beta$. We then extended the
scope to joint ML estimator of two parameters. 

The concepts and the techniques exercised in this paper are believed to
generalize to other signal models, as well as to
joint ML estimation of more than two parameters. The proposed approach 
may therefore serve as a
yardstick for gaining insights and understanding 
concerning the threshold behavior
in more complicated situations, 
including models which are expected to exhibit more
than one threshold with respect to the SNR
(which means
more than one phase transition in the 
analogous physical model). For example, models
of superimposed signals, where each component signal has its own threshold SNR,
or combinations of threshold effects due to non--linearity (as studied here)
with threshold effects that stem from ambiguity. The latter is characteristic,
for example, when the delay of a narrow-band signal is to be estimated
(see, e.g., \cite{Weiss85}).

\section*{Acknowledgment}

Interesting discussions with Yariv Kafri are acknowledged with thanks.



\clearpage
\begin{thebibliography}{AA}

\bibitem{Adler90}
R.~J.~Adler, {\it An Introduction to Continuity, Extrema, and Related Topics
for General Gaussian Processes}, Institute of Mathematical Statistics,
Lecture Notes -- Monograph Series, Vol.\ 12, 1990.

\bibitem{BL73}
I.~F.~Blake and W.~C.~Lindsey, ``Level--crossing problems for random 
processes,''  {\it IEEE Trans.\ Inform.\
Theory}, vol.\ IT--19, no.\ 3, pp.\ 295--315, May 1973.

\bibitem{Derrida80}
B.~Derrida, ``Random--energy model: limit of a family of disordered models,''
{\it Phys.\ Rev.\ Lett.}, vol.\ 45, no.\ 2, pp.\ 79--82, July 1980.

\bibitem{Derrida80b}
B.~Derrida, ``The random energy model,'' {\it Physics Reports} (Review Section
of Physics Letters), vol.\ 67, no.\ 1, pp.\ 29--35, 1980.

\bibitem{Derrida81}
B.~Derrida, ``Random--energy model: an exactly solvable model for disordered
systems,'' 
{\it Phys.\ Rev.\ B}, vol.\ 24, no.\ 5, pp.\ 2613--2626, September 1981.

\bibitem{Floor08}
P.~A.~Floor, ``On the theory of Shannon--Kotel'nikov 
mappings  in joint source--channel coding,''
Ph.D.\ dissertation, Norwegian University of Science and Engineering (NTNU),
Trondheim, Norway, May 2008.

\bibitem{FR09}
P.~A.~Floor and T.~A.~Ramstad, ``On the analysis of 
Shannon--Kotel'nikov mappings,''
arXiv:0904.1538v1 [cs.IT] 7 Apr 2009.

\bibitem{GH69}
R.~G.~Gallager and C.~W.~Helstrom, ``A bound on the probability that 
a Gaussian process exceeds a given function,'' {\it IEEE Trans.\ Inform.\
Theory}, vol.\ IT--15, no.\ 1, pp.\ 163--166, January 1969.

\bibitem{Hekland07}
F.~Hekland, ``On the design and analysis of Shannon--Kotel'nikov 
mappings for joint source--channel coding,''
Ph.D.\ dissertation, Norwegian University of Science and Engineering (NTNU),
Trondheim, Norway, May 2007.

\bibitem{MP71}
M.~B.~Marcus and L.~A.~Shepp, ``Sample behavior of Gaussian processes,''
{\it Proc.\ Sixth Berkeley Symp.\ Math.\ Statist.\ Prob.}, vol.\ 2, pp.\
423--442, 1971.

\bibitem{Merhav08}
N.~Merhav, ``Relations between random coding exponents and
the statistical physics of random codes,''
{\it IEEE Trans.\ Inform.\ Theory}, vol.\ 55, no.\ 1, pp.\ 83--92, January
2009.

\bibitem{MGS10}
N.~Merhav, D.~Guo, and S.~Shamai (Shitz),
``Statistical physics of signal estimation in Gaussian noise:
theory and examples of phase transitions,''
{\it IEEE Trans.\ Inform.\ Theory}, vol.\ 56, no.\ 3, pp.\ 1400--1416, March
2010.

\bibitem{MM09}
M.~M\'ezard and
A.~Montanari, {\it Information, Physics and Computation},
Oxford University Press, 2009.

\bibitem{Ramstad02}
T.~A.~Ramstad, ``Shannon mappings for robust communication,'' 
{\it Telektronikk}, vol.\ 98, no.\ 1, pp.\ 114--128, 2002.

\bibitem{Shepp66}
L.~A.~Shepp, ``Radon--Nykodim derivatives of Gaussian measures,''
{\it Ann.\ Math.\ Statist.}, vol.\ 37, pp.\ 321--354, April 1966.

\bibitem{Slepian62}
D.~Slepian, ``The one--sided barrier problem for Gaussian noise,''
{\it Bell Systems Technical Journal}, vol.\ 41, pp.\ 463--501, March 1962.

\bibitem{Weiss85}
A.~J.~Weiss, {\it Fundamental Bounds in Parameter Estimation},
Ph.D.\ dissertation, Tel Aviv University, Tel Aviv, Israel, June 1985.

\bibitem{WW85}
A. J. Weiss and E. Weinstein, ``A lower bound on the mean square error
in random parameter estimation,''
{\em IEEE Transactions on Information Theory\/},
vol.~IT--31, no.~5, pp.~680--682, September 1985.

\bibitem{WJ65}
J.~M.~Wozencraft and I.~M.~Jacobs, {\it Principles of Communication
Engineering}, John Wiley \& Sons, 1965. Reissued by Waveland Press, 1990.

\bibitem{ZZ69}
M.~Zakai and J.~Ziv, ``On the threshold effect in radar range 
estimation,'' {\it IEEE Trans.\ Inform.\ Theory}, vol.\ IT--15, pp.\ 167--170,
January 1969.

\end{thebibliography}
\end{document}